\documentclass{article}
\usepackage[utf8]{inputenc}
\usepackage[
backend=biber,
style=phys,
sorting=none
]{biblatex}
\usepackage{titling}
\newcommand{\subtitle}[1]{%
  \posttitle{%
    \par\end{center}
    \begin{center}\large#1\end{center}
    \vskip0.5em}%
}

\usepackage{amssymb}
\usepackage{authblk}
\usepackage{tabularx}
\usepackage{physics}
\usepackage{graphicx}
\usepackage[left=1in, right=1in, top=1in, bottom = 1in]{geometry}

\title{Parametric Amplifier Matching Using Legendre Prototypes}
\author[1,2]{Ryan Kaufman} \author[1]{Ofer Naaman}
\affil[1]{Google Quantum AI, Goleta, CA 93117}
\affil[2]{Department of Physics and Astronomy, University of Pittsburgh, Pittsburgh, PA 15260}
\date{\today}

\begin{document}

\maketitle

\section{Introduction}
Matching networks for active, negative-resistance loads can be used to design Josephson parametric amplifiers (JPA) having specified gain and bandwidth characteristics. Chebyshev JPA matching networks were previously described in \cite{naaman2022synthesis} and  \cite{getsinger1963prototypes}, as well as networks based on Butterworth prototypes \cite{ naaman2022synthesis, henoch1963new, roy2015broadband}. A method for calculating prototype coefficients was outlined in \cite{naaman2022synthesis}, Appendix E, including coefficient tables for Chebyshev and Butterworth prototypes.

In this note we describe JPA matching networks based on Legendre polynomials \cite{Chryssomallis1999FilterPolynomials}. These networks typically exhibit lower ripple and gentler roll-off than Chebyshev networks with similar parameters, and can be viewed as bridging the gap between Butterworth and Chebyshev ones. We tabulate prototype coefficients for parametric amplifiers based on Legendre polynomials with a range of gain and ripple parameters, and for a range of network orders. We also use this opportunity to further illustrate the synthesis of these networks based on Ref.~\cite{naaman2022synthesis}. 

\section{Legendre Filter Characteristics}
\begin{figure}
    \centering
    \includegraphics[width=4in]{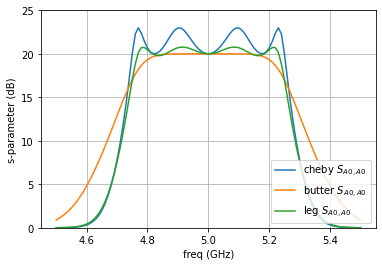}
    \caption{Comparison of gain profiles for JPA designed using Chebyshev (blue), Butterworth (orange), and Legendre (green) polynomials. All networks designed for 20 dB gain at the center frequency, and a  3 dB points bandwidth of 500 MHz.}
    \label{fig:compare_gains}
\end{figure}

Figure~\ref{fig:compare_gains} shows calculated gain vs frequency for JPAs designed with fourth-order Chebyshev, Butterworth, and Legendre polynomials. All networks were designed for a gain of 20 dB and a 3 dB bandwidth of 500 MHz. While the bandwidth of a Butterworth network is always defined at the 3 dB points, in Chebyshev networks this is only the case for 3 dB ripple designs, i.e., the bandwidth in Chebyshev networks is specified in relation to the ripple. The Legendre networks also specify a ``ripple" parameter, but in this case it only controls the gain at the cutoff frequency and not the actual gain ripple in the band. The Chebyshev and Legendre designs in Fig.~\ref{fig:compare_gains} both specify a ripple parameter of 3 dB.

The coupled-modes formalism in \cite{naaman2022synthesis} describes the strength of the parametric pump via a parameter $\beta$. This is the parametric coupling rate normalized to $2\gamma_0$, where $\gamma_0$ is the geometric mean of all dissipation rates at the ports of the device. Comparing the designs in Fig.~\ref{fig:compare_gains}, we find that $\beta_\mathrm{chebyshev}=0.342$, $\beta_\mathrm{legendre}=0.285$, and $\beta_\mathrm{butter}=0.179$, meaning that, with all other parameters held the same, the Legendre JPA requires higher pump coupling than Butterworth, but lower pump coupling than Chebyshev.

The figure also illustrates the main differences between the various response types. We see that Legendre networks have sharper roll-off than Butterworth ones, but not as sharp as Chebyshev networks. We also see that the ripple in the band is lower in Legendre networks compared with that of Chebyshev network, while for Butterworth networks the pass band is flat.

\section{Power Loss Function and 4-Pole Synthesis}\label{sec:coefficients}

We will follow a matching network synthesis flow based on the insertion-loss method. We will go through constructing a power loss function for a Legendre JPA, and then apply the procedure outlined in \cite{naaman2022synthesis} Appendix E to the specific case of a 4-pole, 20 dB amplifier with 0.5 dB ripple.

\subsection{Power Loss Function}
The power loss function is defined as
\begin{equation}\label{eq:pil_general}
    P_{IL}=A\left[1+k^2P^2_N(\omega)\right]
\end{equation}
where $A$ and $k$ are constants and $P_N(\omega)$ is the Legendre polynomial of order $N$ as a function of frequency $\omega$ normalized to the cutoff frequency $\omega_c=2\pi\times1$~Hz. The power loss function is related to the reflection coefficient of the network via
\begin{equation}\label{eq:gamma_sq_from_pil}
    |\Gamma(\omega)|^2=1-\frac{1}{P_{IL}}
\end{equation}

For a design with a signal power gain $G$ (in linear power units) at the center of the band, we require \cite{naaman2022synthesis}
\begin{equation}\label{eq:gain_center}
    |\Gamma(0)|^2 = \frac{1}{4G-2} \equiv \tilde{G}^{-1}.
\end{equation}
Further, for a specified ripple parameter $R^{dB}$ in dB units, we calculate the linear power gain $G_m$ at the cutoff frequency using $10\log_{10}G_m = G^{dB}-R^{dB}$, where $G^{dB}$ is the signal gain in dB units. We therefore require that at the cutoff frequency
\begin{equation}\label{eq:gain_edge}
    |\Gamma(1)|^2 = \frac{1}{4G_m-2}\equiv \tilde{G}_m^{-1}.
\end{equation}
In both Eqs.~(\ref{eq:gain_center}) and~(\ref{eq:gain_edge}) we mark the `double sideband' gain quantities with a tilde.

Using the conditions Eqs.~(\ref{eq:gain_center}) and~(\ref{eq:gain_edge}) together with Eqs.~(\ref{eq:gamma_sq_from_pil}) and~(\ref{eq:pil_general}), we get two equations for $A$ and $k^2$
\begin{align}
    k^2 &= \frac{\tilde{G}^{-1}-\tilde{G}_m^{-1}}{P_N^2(0)(1-\tilde{G}^{-1})+\tilde{G}_m^{-1}-1}, \label{eq:k2}\\
    A &= \frac{1}{(1+k^2)(1-\tilde{G}_m^{-1})} \label{eq:pil0}.
\end{align}
Note that Eq.~(\ref{eq:k2}) explicitly contains $P^2_N(0)$. This is because unlike Chebyshev polynomials, $P_N(0)$ is different for different orders. While $P_N(0)=0$ for odd $N$, it is finite for even $N$, and its value depends on $N$. At the cutoff frequency, we have used $P_N(1)=1$ for all $N$. Eqs.~(\ref{eq:k2}) and~(\ref{eq:pil0}) can be simplified further for odd-$N$ Legendre networks. We now have both $A$ and $k^2$ specified in terms of the order of the network, the required gain, and the ripple parameter, and we can write down the power loss function of Eq.~(\ref{eq:pil_general}) as a function of frequency. 

\subsection{Finding Coefficients for a Fourth-order Network}
To illustrate the synthesis flow, we will step through the calculations for extracting the prototype coefficients for a fourth-order Legendre network, matching a JPA for a specified gain of 20 dB with a ripple parameter of 0.5 dB. Note that we do not necessarily recommend the use of 4th order networks, as in most cases 3rd order is sufficient; we do so here just for illustration purposes.

From the gain, ripple, and order, we can write down the power loss function Eq.~(\ref{eq:pil_general}), and from it the reflection coefficient $|\Gamma(s)|^2$, where we have changed variable to $s=j\omega$.
\begin{equation}
    |\Gamma(s)|^2 = \frac{P_{IL}-1}{P_{IL}}=\frac{0.0069s^8+0.01183s^6+0.006256s^4+0.001s^2+0.003}{0.0069s^8+0.01183s^6+0.006256s^4+0.001s^2+1.003}.
\end{equation}
Next we need to factor $|\Gamma(s)|^2=\Gamma(s)\Gamma^*(s)$. We first find the complex roots of the $8^\mathrm{th}$-order polynomials in the numerator and denominator, and then reconstruct $\Gamma(s)$ as a ratio of $4^\mathrm{th}$-order polynomials from a subset of these roots. 

\begin{figure}
    \centering
    \includegraphics[width=2.4in]{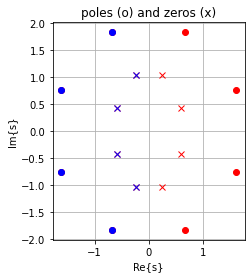}
    \caption{Complex poles and zeros of $|\Gamma(s)|^2$ for a fourth-order Legendre amplifier matching network. Out of the eight roots of numerator (zeros) and denominator (poles), we pick the four that are in the left-half-plane are in blue.}
    \label{fig:poles_zeros}
\end{figure}

The complex roots of the numerator of $|\Gamma(s)|^2$, which we will call `zeros', $z_i$, are:
\begin{align*}
    z_{1,2} &= 0.2424 \pm 1.0395j \\
    z_{3,4} &= -0.2424 \pm 1.0395j \\
    z_{5,6} &= 0.5894 \pm 0.4275j \\
    z_{7,8} &= -0.5894 \pm 0.4275j
\end{align*}
and the roots of the denominator of $|\Gamma(s)|^2$, which we call `poles', $p_i$, are:
\begin{align*}
    p_{1,2} &= 1.6152 \pm 0.7568j \\
    p_{3,4} &= -1.6152 \pm 0.7568j \\
    p_{5,6} &= 0.6688 \pm 1.8277j \\
    p_{7,8} &= -0.6688 \pm 1.8277j. 
\end{align*}
Fig.~\ref{fig:poles_zeros} shows the location of the zeros (crosses) and poles (circles) on the complex s-plane. We now choose four out of the eight roots of each of the polynomials, and we pick the ones that are in the left half of the complex s-plane, as shown in blue in Fig.~\ref{fig:poles_zeros}. These are $z_3,\;z_4,\;z_7,\;z_8$ from the numerator, and $p_3,\;p_4,\;p_7,\;p_8$ from the denominator. 

From the selected roots, we can construct the polynomials $R$ and $D$ such that $\Gamma(s)=R/D$,
\begin{align}
    R(s) &= (s-z_3)(s-z_4)(s-z_7)(s-z_8) \\
    D(s) &= (s-p_3)(s-p_4)(s-p_7)(s-p_8)
\end{align}
and from $\Gamma(s)$ we can calculate the input impedance of the network, expanding $D$ and $R$: 
\begin{equation}\label{eq:z_polynomials}
    Z(s)=\frac{1+\Gamma(s)}{1-\Gamma(s)}=\frac{D(s)+R(s)}{D(s)-R(s)}=\frac{2.0s^4+6.262s^3+13.53s^2+19.09s+12.65}{2.904s^3+9.049s^2+14.89s+11.45}. 
\end{equation}
We see that the numerator of the impedance $Z(s)$ is a fourth order polynomial, and the denominator is a third-order polynomial. We can now perform polynomial long division to express $Z(s)$ as a continued fraction \cite{naaman2022synthesis}. 
\begin{equation}
    Z(s)=0.6886s+\cfrac{1}{0.8864s+\cfrac{1}{0.8918s+\cfrac{1}{0.2903s+\cfrac{1}{1.1055}}}}.
    \label{eq:continued_fraction}
\end{equation}
The coefficients of the prototype for a fourth-order Legendre matching network with the specified gain and ripple are therefore the coefficients of the continued-fraction expansion
\begin{equation}\label{eq:coeffs}
    g_i=\left\{1.0,\;0.6886,\;0.8864,\;0.8918,\;0.2903,\;1.1055\right\},
\end{equation}
and we have added $g_0=1.0$ to the list, representing the conductance of the source.

Following \cite{naaman2022synthesis}, we can now use the coefficients Eq.~(\ref{eq:coeffs}) to find the  port coupling $\gamma_0$, mode couplings $\beta_{k,k+1}$, and pump coupling strength $\beta_p$ for the coupled-mode system representing the amplifier,
\begin{align}
    \gamma_0 = \gamma_{45} &= \frac{\Delta\omega}{g_4g_5}, \\
    \beta_{k,k+1} &= \frac{\Delta\omega}{2\gamma_0\sqrt{g_kg_{k+1}}},\;\;\;k=1\dots3,\\
    \beta_p &= \frac{g_4g_5}{2g_0g_1},
\end{align}
where $\Delta\omega$ is the full bandwidth of the network.

\begin{figure}
    \centering
    \includegraphics[width=3.2in]{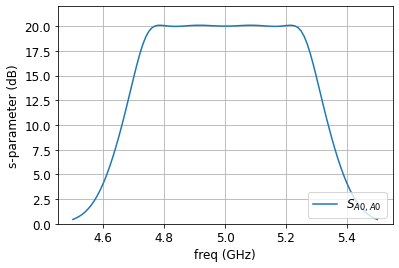}
    \caption{Gain characteristics of a $4^\mathrm{th}$-order Legendre JPA calculated with the coefficients in Eq.~(\ref{eq:coeffs}), with bandwidth of 500 MHz and center frequency of 5 GHz.}
    \label{fig:leg4}
\end{figure}

Figure~\ref{fig:leg4} shows the calculated gain profile \cite{naaman2022synthesis} of an amplifier designed with the coefficients found in Eq.~(\ref{eq:coeffs}), with a center frequency of 5 GHz and a bandwidth of 500 MHz. We find $\gamma_0/2\pi=1.558$~GHz, $\beta_{12}=0.315$, $\beta_{23}=0.18$, $\beta_{34}=0.205$, and $\beta_p=0.233$. Table~\ref{tab:leg_amp_coefs} lists prototype coefficients for second-, third-, and fourth-order Legendre JPA networks for several gain and ripple parameters.

\section{Calculating Circuit Parameters}
Following the procedure in \cite{naaman2022synthesis}, we can use the filter prototype coefficients given in Table~\ref{tab:leg_amp_coefs} to calculate device component values based on a coupled-resonator network topology. The procedure requires the following inputs: the order of the filter, $N$, the prototype coefficients $g_0,\dots,g_{N+1}$, the characteristic impedance $Z_k$ of the $N$ resonators, the fractional bandwidth of the network $w$, and the center frequency of the network $\omega_0$. We will continue with the example in Sec.~\ref{sec:coefficients} to design a 4th order Legendre network with $\omega_0/2\pi = 5$~GHz, a fractional bandwidth of $w=0.1$, gain of 20~dB, ripple of 0.5~dB, and the coefficients in Eq.~(\ref{eq:coeffs}). 

The schematic of the synthesized degenerate amplifier circuit is shown in Fig.~\ref{fig:circuit_synthesis} as a 1-port network. Resonator $R_1$ on the right contains the nonlinear, parametrically pumped inductance $L_\mathrm{snake}$, and the rest of the resonators are passive and constitute the matching network. We label the resonators starting from the parametric element on the right, to align with labeling of the prototype coefficients that regard the parametric element as the source. The nonlinear element we consider here is the `snake', an rf-SQUID array used in Ref.~\cite{white2022readout}, but from a network design point of view, this could equivalently be a standard dc-SQUID \cite{mutus2013design}. A simulation using Keysight ADS is shown in Fig.~\ref{fig:ads}.

\subsection{Resonator Impedances}
\begin{figure}[ht]
    \centering
    \includegraphics[width=\textwidth]{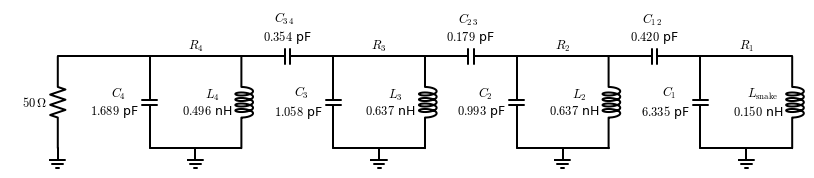}
    \caption{Circuit schematic of a 4th order Legendre matched degenerate parametric amplifier.}
    \label{fig:circuit_synthesis}
\end{figure}

\begin{figure}[hb]
    \centering
    \includegraphics[width=3.2in]{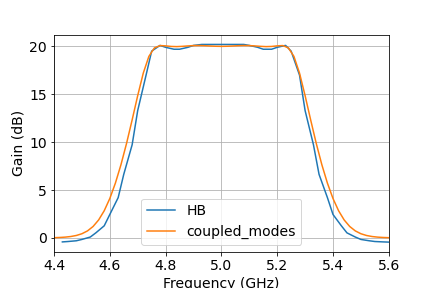}
    \caption{Comparison of the gain calculated in Sec.~\ref{sec:coefficients} using the inverse of the coupled-mode matrix (orange), and the gain calculated from harmonic balance simulation in Keysight ADS using the circuit and parameters in Fig.~\ref{fig:circuit_synthesis}.}
    \label{fig:ads}
\end{figure}

The first step in calculating circuit component values is to determine the resonator impedances. We first work on resonator $R_1$ in Fig.~\ref{fig:circuit_synthesis}, which contains the capacitively shunted Josephson `snake' element $L_\mathrm{snake}$ \cite{white2022readout}. We target the snake's operating flux bias point to give $L_\mathrm{snake}=150$~pH, so that the impedance of resonator $R_1$ is $Z_1=\omega_0L_\mathrm{snake}=4.7\,\Omega$.

Next, we work on the last resonator, $R_4$. Normally, the impedance of this resonator can be rather arbitrary, since we could use an admittance inverter $J_{45}$, implemented for example as a coupling capacitor, to set the correct coupling strength to the $50\,\Omega$ port \cite{naaman2022synthesis}. However, we would like to avoid an input coupling capacitor because it invariably adds its own, often unacceptable, frequency dependence of the coupling strength. One approach that we will take here, is to constrain the impedance of $R_4$ to give the correct coupling strength to the environment \cite{naaman2019high}, thus obviating the need to use a coupling capacitor (essentially, making $J_{45}$ trivial). We do so by requiring that $J_{45}Z_0=1$, where
\begin{equation}   
    J_{45}=\sqrt{\frac{w}{g_4g_5Z_4Z_0}}.
\end{equation}
This leads to $R_4$ impedance being set to
\begin{equation}
    Z_4 = \frac{wZ_0}{g_4g_5}=15.58\,\Omega.
\end{equation}
The rest of the resonator impedances can be now chosen, and we chose $Z_2=Z_3=20\,\Omega$. 

\subsection{Coupling Capacitors}
Now that the resonator impedances are set, $Z_1=4.7\,\Omega$, $Z_2=Z_3=20\,\Omega$, and $Z_4=15.58\,\Omega$, we can calculate the rest of the admittance inverter values, and from those, the coupling capacitors. 

Inverter $J_{01}$ will be taken care of by setting the correct pump amplitude, so that the idler presents an admittance of $-R^{-1}_{PA}$, so that 
\begin{equation}
    \frac{R_{PA}}{Z_1} = \frac{g_0g_1}{w}, 
\end{equation}
and there is nothing further we need to do on the circuit design front with respect to this coupling. Inverter $J_{45}$ is likewise already taken care of by choosing the impedance $Z_4$ above.

We are left to determine the values of the rest of the admittance inverters, which we can calculate via
\begin{equation}
  J_{jk} = w\sqrt{\frac{1}{g_j g_k Z_j Z_k}}.  
\end{equation}
From the admittance inverter values we can calculate the coupling capacitors using 
\begin{equation}
    C_{jk} = \frac{J_{jk}}{\omega_0}.
\end{equation}
In our design example, we get:
\begin{align}
   C_{12} &= \frac{w}{\omega_0}\sqrt{\frac{1}{g_1g_2Z_1Z_2}} = 0.420\,\mathrm{pF}, \\
   C_{23} &= \frac{w}{\omega_0}\sqrt{\frac{1}{g_2g_3Z_2Z_3}} = 0.179\,\mathrm{pF}, \\
   C_{34} &= \frac{w}{\omega_0}\sqrt{\frac{1}{g_3g_4Z_3Z_4}} = 0.354\,\mathrm{pF}.
\end{align}

\subsection{Resonator Components}
Now that the coupling capacitors are calculated, we can continue to construct the resonators in the network, implemented as shunt lumped-element parallel $LC$ resonators. In each resonator, the capacitor will absorb the negative capacitance associated with its neighboring admittance inverters.
\begin{align}
    C_1 &= \frac{1}{\omega_0Z_1}-C_{12} = 6.352\,\mathrm{pF}, \\
    C_2 &= \frac{1}{\omega_0Z_2}-C_{12}-C_{23} = 0.993\,\mathrm{pF}, \\
    C_3 &= \frac{1}{\omega_0Z_3}-C_{23}-C_{34} = 1.058\,\mathrm{pF}, \\
    C_4 &= \frac{1}{\omega_0Z_4}-C_{34} = 1.680\,\mathrm{pF}.
\end{align}
Lastly, we can calculate the inductors easily with $L_i = Z_i/\omega_0$, so that $L_2=L_3=0.637$~nH, and $L_4=0.496$~nH. Recall that we already set $L_1=L_\mathrm{snake}=150$~pH from the outset, by setting the flux bias operating point of the amplifier. This completes the design as shown in Fig.~\ref{fig:circuit_synthesis}.

\begin{table*}[h!]
\caption{Legendre coefficients for several values of amplifier gain and gain ripple, and for 2-, 3-, and 4-section networks, using the power loss function Eq.~(\ref{eq:pil_general}). Coefficient $g_0$ refers to the active load.\label{tab:leg_amp_coefs}}
\begin{tabularx}{\textwidth}{
|l||l|>{\centering\arraybackslash}X|>{\centering\arraybackslash}X|>{\centering\arraybackslash}X|>{\centering\arraybackslash}X|>{\centering\arraybackslash}X|>{\centering\arraybackslash}X|>{\centering\arraybackslash}X|
}
\hline
\textrm{Signal gain}&
\textrm{ripple}&
\textrm{order}&
\multicolumn{1}{c|}{\textrm{$g_0$}}&
\multicolumn{1}{c|}{\textrm{$g_1$}}&
\multicolumn{1}{c|}{\textrm{$g_2$}}&
\multicolumn{1}{c|}{\textrm{$g_3$}}&
\multicolumn{1}{c|}{\textrm{$g_4$}}&
\multicolumn{1}{c|}{\textrm{$g_5$}}\\
\hline

$G=17$~dB& $R=0.1$~dB & 2 &  1.0  &  0.2619 & 0.1348 & 1.1528 & &\\
&& 3 &  1.0  &  0.4924 & 0.4649 & 0.2567 & 0.8674 &\\
&& 4 &  1.0  &  0.6884 & 0.7607 & 0.7975 & 0.2304 & 1.1528\\

& $R=0.5$~dB & 2 &  1.0  &  0.3898 & 0.2075 & 1.1528 & &\\
&& 3 &  1.0  &  0.6304 & 0.6100 & 0.3499 & 0.8674 &\\
&& 4 &  1.0  &  0.8099 & 0.9137 & 0.9939 & 0.2996 & 1.1528\\

& $R=1.0$~dB &2 &  1.0  &  0.4640 & 0.2542 & 1.1528 & &\\
&& 3 &  1.0  &  0.7020 & 0.6886 & 0.4060 & 0.8674 &\\
&& 4 &  1.0  &  0.8661 & 0.9893 & 1.0989 & 0.3413 & 1.1528\\

& $R=3.0$~dB &2  &  1.0  &  0.6181 & 0.3762 & 1.1528 & &\\
&& 3 &  1.0  &  0.8459 & 0.8477 & 0.5396 & 0.8674 &\\
&& 4 &  1.0  &  0.9598 & 1.1333 & 1.3121 & 0.4440 & 1.1528\\
\hline

$G=20$~dB & $R=0.1$~dB & 2 &  1.0  &  0.2080 & 0.1218 & 1.1055 & &\\
&& 3 &  1.0  &  0.4084 & 0.4399 & 0.2250 & 0.9045 &\\
&& 4 &  1.0  &  0.5833 & 0.7364 & 0.7160 & 0.2244 & 1.1055\\

& $R=0.5$~dB & 2 &  1.0  &  0.3105 & 0.1868 & 1.1055 & &\\
&& 3 &  1.0  &  0.5244 & 0.5778 & 0.3055 & 0.9045 &\\
&& 4 &  1.0  &  0.6886 & 0.8864 & 0.8918 & 0.2903 & 1.1055\\

& $R=1.0$~dB & 2 &  1.0  &  0.3703 & 0.2283 & 1.1055 & &\\
&& 3 &  1.0  &  0.5849 & 0.6527 & 0.3534 & 0.9045 &\\
&& 4 &  1.0  &  0.7379 & 0.9608 & 0.9860 & 0.3295 & 1.1055\\

& $R=3.0$~dB & 2 &  1.0  &  0.4972 & 0.3343 & 1.1055 & &\\
&& 3 &  1.0  &  0.7066 & 0.8061 & 0.4656 & 0.9045 &\\
&& 4 &  1.0  &  0.8219 & 1.1025 & 1.1789 & 0.4239 & 1.1055\\
\hline

$G=25$~dB & $R=0.1$~dB & 2 &  1.0  &  0.1455 & 0.0995 & 1.0579 & &\\
&& 3 &  1.0  &  0.3074 & 0.3892 & 0.1849 & 0.9453 &\\
&& 4 &  1.0  &  0.4552 & 0.6767 & 0.6122 & 0.2094 & 1.0579\\

& $R=0.5$~dB & 2 &  1.0  &  0.2179 & 0.1522 & 1.0579 & &\\
&& 3 &  1.0  &  0.3964 & 0.5117 & 0.2497 & 0.9453 &\\
&& 4 &  1.0  &  0.5401 & 0.8170 & 0.7619 & 0.2688 & 1.0579\\

& $R=1.0$~dB & 2 &  1.0  &  0.2607 & 0.1853 & 1.0579 & &\\
&& 3 &  1.0  &  0.4432 & 0.5786 & 0.2877 & 0.9453 &\\
&& 4 &  1.0  &  0.5806 & 0.8872 & 0.8422 & 0.3035 & 1.0579\\

& $R=3.0$~dB & 2 &  1.0  &  0.3537 & 0.2681 & 1.0579 & &\\
&& 3 &  1.0  &  0.5380 & 0.7176 & 0.3750 & 0.9453 &\\
&& 4 &  1.0  &  0.6516 & 1.0215 & 1.0083 & 0.3849 & 1.0579\\
\hline

$G=30$~dB & $R=0.1$~dB & 2 &  1.0  &  0.1040 & 0.0792 & 1.0321 & &\\
&& 3 &  1.0  &  0.2369 & 0.3357 & 0.1543 & 0.9689 &\\
&& 4 &  1.0  &  0.3639 & 0.6065 & 0.5316 & 0.1912 & 1.0321\\

& $R=0.5$~dB & 2 &  1.0  &  0.1562 & 0.1207 & 1.0321 & &\\
&& 3 &  1.0  &  0.3066 & 0.4416 & 0.2074 & 0.9689 &\\
&& 4 &  1.0  &  0.4338 & 0.7343 & 0.6610 & 0.2439 & 1.0321\\

& $R=1.0$~dB & 2 &  1.0  &  0.1872 & 0.1466 & 1.0321 & &\\
&& 3 &  1.0  &  0.3435 & 0.4998 & 0.2382 & 0.9689 &\\
&& 4 &  1.0  &  0.4675 & 0.7987 & 0.7305 & 0.2743 & 1.0321\\

& $R=3.0$~dB & 2 &  1.0  &  0.2560 & 0.2103 & 1.0321 & &\\
&& 3 &  1.0  &  0.4189 & 0.6216 & 0.3081 & 0.9689 &\\
&& 4 &  1.0  &  0.5282 & 0.9227 & 0.8748 & 0.3441 & 1.0321\\
\hline
\end{tabularx}
\end{table*}

\printbibliography
\end{document}